\documentclass[prd,preprint,nofootinbib,amsmath,amssymb,superscriptaddress]{revtex4-1}
\pdfoutput=1
\usepackage{braket}
\usepackage[dvipdfmx]{graphicx}
\usepackage{dcolumn}
\usepackage{bm}
\usepackage[colorlinks=true,link color=blue,cite color=blue,urlcolor=black]{hyperref}

\begin{document}

\preprint{ICRR-Report-662-2013-11, IPMU-13-204}
\vspace*{2cm}
\title{\Large I-ball formation with logarithmic potential}

\author{Masahiro Kawasaki}
 \email{kawasaki@icrr.u-tokyo.ac.jp}
 \affiliation{%
 Institute for Cosmic Ray Research, University of Tokyo, \\
 Kashiwa, Chiba, 277-8582, Japan}
\affiliation{ Kavli Institute for the Physics and Mathematics of the Universe(WPI),\\
Todai Insititute for Advanced Sturdy, The University of Tokyo, 
Kashiwa, Chiba 277-8568, Japan
}%
 \author{Naoyuki Takeda}%
 \email{takedan@icrr.u-tokyo.ac.jp}
\affiliation{%
 Institute for Cosmic Ray Research, University of Tokyo, \\
 Kashiwa, Chiba, 277-8582, Japan}
\date{\today}

\begin{abstract}
A coherently oscillating real scalar field with potential shallower than 
quadratic one fragments into spherical objects called I-balls. 
We study the I-ball formation for logarithmic potential which appears in many
cosmological models. 
We perform lattice simulations and find that the I-balls are formed 
when the potential becomes dominated 
by the quadratic term. 
Furthermore, we estimate the I-ball profile assuming that the adiabatic invariant is 
conserved during formation and obtain the result that agrees to the numerical simulations. 
\end{abstract}

\pacs{Valid PACS appear here}
\maketitle



\section{Introduction}
\label{sec:intro}

It is well known that scalar fields play important roles in cosmology.
The most famous example is the inflaton field which causes the accelerated 
cosmic expansion ($=$ inflation) and generates density perturbations as observed by
CMB observations. 
Some scalar fields existing in the early universe form (quasi-)stable localized 
objects or solitons whose stability is ensured by some conserved quantity such as
a topological number and a global $U(1)$ charge.  
Topological numbers lead to formation of monopoles, cosmic strings 
and domain walls and hence they are called topological 
defects~\cite{Kibble:1976sj}.
On the other hand, a scalar field with global $U(1)$ symmetry forms a spherical
object called Q~ball~\cite{Coleman:1985ki}.
The formation of these solitons would significantly affect the 	
cosmological scenario. 
For example, domain walls, if formed through spontaneous breaking of a discrete 
symmetry, soon dominate the universe and cause a serious cosmological
difficulty~\cite{Zeldovich:1974uw}. 
In another example, when the Affleck-Dine field (which is a flat direction 
in the scalar potential of the minimal supersymmetric(SUSY) standard model) 
fragments into Q-balls, its existence significantly changes the scenario 
of the Affleck-Dine baryogenesis~\cite{Kusenko:1997si,Enqvist:1997si,Enqvist:1998en,Kasuya:1999wu,Kasuya:2000wx,Kasuya:2001hg}. 

Thus, the conserved quantities seem crucial for 
formation of stable solitons. 
However, even though there is no evident charge, in some potentials, 
the scalar field fragments into long lived and localized spherical 
objects~\cite{Bogolyubsky:1976nx}. 
In this paper, we call this object  ``I-ball" following Ref.~\cite{Kasuya:2002zs}.\footnote{%
The name ``oscillon" is also used in the literature.}
I-balls are formed through coherent oscillations of a 
real scalar field $\phi$ with potential shallower than quadratic~\cite{Amin:2013ika}.\footnote{%
More precise necessary condition for the formation of the I-ball is given in~\cite{Amin:2013ika} as follows.
For the model described by
${\cal L}=X+\xi_2 X^2-(1/2\phi^2+\lambda_3/3\phi^3+\lambda_4/4\phi^4)$ where 
$X\equiv-1/2\partial_{\mu}\phi\partial^{\mu}\phi$, 
the condition is given as $\xi_2-\lambda_4+10/9\lambda_3^2>0$.
In the case that the kinetic term is canonical, the condition is given by setting $\xi_2=0$ as
$-\lambda_4+10/9\lambda_3^2>0$.
}
Ref.~\cite{Kasuya:2002zs} pointed out that the oscillating real scalar field has 
an adiabatic invariant which plays a role of the conserved charge 
and accounts for the stability of the I-ball. 
However, the adiabatic invariance requires that the scalar potential $V(\phi)$ 
should be dominated by the quadratic term $m^2 \phi^2$.
Therefore, it is not certain whether soliton-like objects are formed for other type
of scalar potentials. 
Furthermore, because of the complexity of the non-perturbative evolution of the 
fluctuations of the scalar field, the I-ball formation and its stability is not 
sufficiently understood. 

The slightly shallower potential than quadratic one attracts the attention in the various situations in the early Universe. 
For instance, the detection of the gravitational waves by BICEP2~\cite{Ade:2014xna} which gives 
strong evidence for the realization of the inflation, prefers the nearly quadratic inflation as
$V\propto\phi^n(n=2.0^{+0.9}_{-0.8})$~\cite{Ellis:2014rxa}.
In the case that the potential of the inflaton is shallower than quadratic as like $n<2$,
it was shown that oscillation of the inflaton during the reheating epoch leads to the formation of I-balls 
in the previous studies~\cite{Amin:2011hj,Amin:2010xe}. 
In those cases, the I-ball formation after inflation has interesting cosmological 
effects such as enhancement of the inflaton 
decay~\cite{McDonald:2001iv,Hertzberg:2010yz} and production of gravitational 
waves~\cite{Zhou:2013tsa}. 

Moreover, the potentials of the inflaton and other scalar fields can be shallower
by correction due to interactions with other fields. 
For instances, considering the supersymmetric (SUSY) theory, when the SUSY breaking is established by  
the gauge mediation~\cite{de Gouvea:1997tn}, the potential of the scalar field has the logarithmic potential, 
which is shallower than the quadratic potential. 
In this case the scalar field has a $U(1)$ charge and Q balls are formed.
However, if the phase direction has a large mass for some reason and the motion 
of the scalar field is restricted in the radial direction, I-balls may be formed. 
Furthermore, this logarithmic potential appears in more general situations, especially during the reheating epoch. 
During the reheating epoch, decay products of the inflaton or of other fields make the thermal bath and then this 
thermal bath gives the thermal correction to the potential of the scalar field as logarithmic potential $\sim T^4\log(\phi^2/T^2)$~\cite{Mukaida:2012qn}. 
The formation of the I-ball due to this thermal logarithmic potential would change the decay process of various models of the cosmology  
as like curvaton scenario~\cite{Enqvist:2001zp,Lyth:2001nq,Moroi:2001ct},  non-thermal leptogenesis scenarion~\cite{Murayama:1992ua,Murayama:1993em}. 
Thus, to determine the cosmological scenario correctly, we have to study the possibility of the I-ball formation with logarithmic potential.

In the previous studies~\cite{Amin:2010xe,Amin:2011hj}, 
the formation of the I-ball for the shallower potential has been confirmed. 
In \cite{Amin:2010xe}, the potentials are quasi-quadratic potentials where the polynomial terms are 
added to the quadratic potential and we can prove the conservation of the adiabatic invariant. 
For the case of the logarithimic potential, the potential cannot be written as polynomials. 
If the quadric term of the potential is important 
for the formation of the I-ball, the formation of the I-ball for the logarithmic 
potential is nontrivial. 
\footnote{
For the case that the potential is square root of the field, the formation of the I-ball 
is confirmed in~\cite{Amin:2011hj}.
}
Therefore, in this paper we newly study the possibility of the I-ball formation 
with logarithmic potential in expanding universe~\cite{Farhi:2007wj}. 
As this process of the I-ball formation is dominated by the non-linear 
evolution of the fluctuations of the scalar field, we perform the lattice 
simulation.
Furthermore, we estimate the field configuration of the I-ball analytically 
assuming that the adiabatic invariant is conserved.

The organization to this paper is as follows. At first, we confirm the 
existence of the I-ball executing the lattice simulation in Sec.~\ref{sec:SIMULATION}. 
Secondary, to clarify the dynamics of the I-ball 
formation for the logarithmic potential, we analytically estimate the 
profile of the I-ball configuration in Sec.~\ref{sec:Analytical estimate}. 
Finally, we conclude this paper in Sec.~\ref{sec:CONCLUSION}.


\section{SIMULATION}
\label{sec:SIMULATION}


In this section, we confirm the formation of I-balls with logarithmic potential. 
Since the process is dominated by the non-linear evolution 
of the fluctuations of the scalar field,  we perform the lattice simulations. 
The equation of motion is integrated by the leapfrog method with $4$th order and 
the spatial derivative is approximated by the Central-Difference formulas with 
$4$th order.
In this paper, in order to follow the scalar field dynamics for sufficiently 
long time with cosmic expansion, we have performed $1+1$-dimensional numerical 
simulations.

We take the following potential for the scalar field $\phi$:
\begin{equation}\label{eq:simulation-1}
   V=M^2\log\left(1+\phi^2\right),
\end{equation}
where $M$ is the typical scale of the system and $\phi$ is the scalar fields. 
The equation of motion for the scalar fields in $1+1$-dimensional expanding 
Universe is given as
\begin{equation}\label{eq:sim-2}
   \ddot{\phi}+H\dot{\phi}-\frac{\nabla^2}{a^2}\phi+V'=0,
\end{equation}
where a dot and a prime are the derivatives with respect to the cosmic time $t$
and the field $\phi$, respectively, $H$ is the Hubble parameter and 
$a$ is the scale factor. 
Here we suppose that the universe expands as 
$a\propto t^{2/3}$ like the matter-dominant case.\footnote{%
The time dependence of the scale factor we have adopted is for the case of 
matter dominated universe in 3-D and hence it is ad hoc in 1-D. 
We include the Hubble expansion in order to see stable I-balls in the simulation. 
Without the cosmic expansion I-balls collides frequently and are destroyed 
in the 1-D simulation. 
The cosmic expansion dilutes the I-balls and avoids unwanted collisions.
So the precise time dependence of the scale factor is not important. 
To confirm this, we have performed the simulations for the scale factor to evolve as 
radiation dominated universe and found the result is not changed.
}

In the numerical simulation, we take the physical variables to be 
dimensionless as
$V/M^2\rightarrow V$, $tM\rightarrow t$  and $xM\rightarrow x$.
The initial conditions are taken as 
\begin{equation}
\begin{split}
   &\phi(x,t=t_0) = \phi_0 +\delta\phi(x),\\
   &\dot{\phi}(x,t=t_0) = 0,\\
   &H(t=t_0) = M,
\end{split}
\end{equation}
where $\delta\phi$ is set by the Gaussian distribution and its variance is 
set to be $10^{-5}$ i.e. probability of the distribution $P(\delta\phi_x)$ is given as
\begin{equation}
P(\delta\phi_x)=\frac{1}{\sqrt{2\pi\sigma^2}}\exp\left(-\frac{\delta\phi^2_x}{2\sigma^2}\right),
\end{equation}
where 
$\sigma=10^{-5}$.
For this distribution, the initial power spectrum is given as 
$\braket{|\delta\phi_k|^2}=\sigma^2L$ where $L$ is the box size of the simulation.

For the initial amplitude of the homogeneous part $\phi_0$, we take 4 different
values as $\phi_0=\{0.3,1,10,100\}$. 
The box size is chosen as $L=(0.05-5)\times H^{-1}(t=t_0)$ 
and the number of grids is large enough to resolve the size of I-balls as 
$N_{\rm grid}={128,256,512,1024,2048}$. 
The time step $dt$ is set to be $dt/dx< 1/3$. 

\begin{figure}[htbp]
\begin{center}
\begin{tabular}{c c}
\resizebox{80mm}{!}{\includegraphics{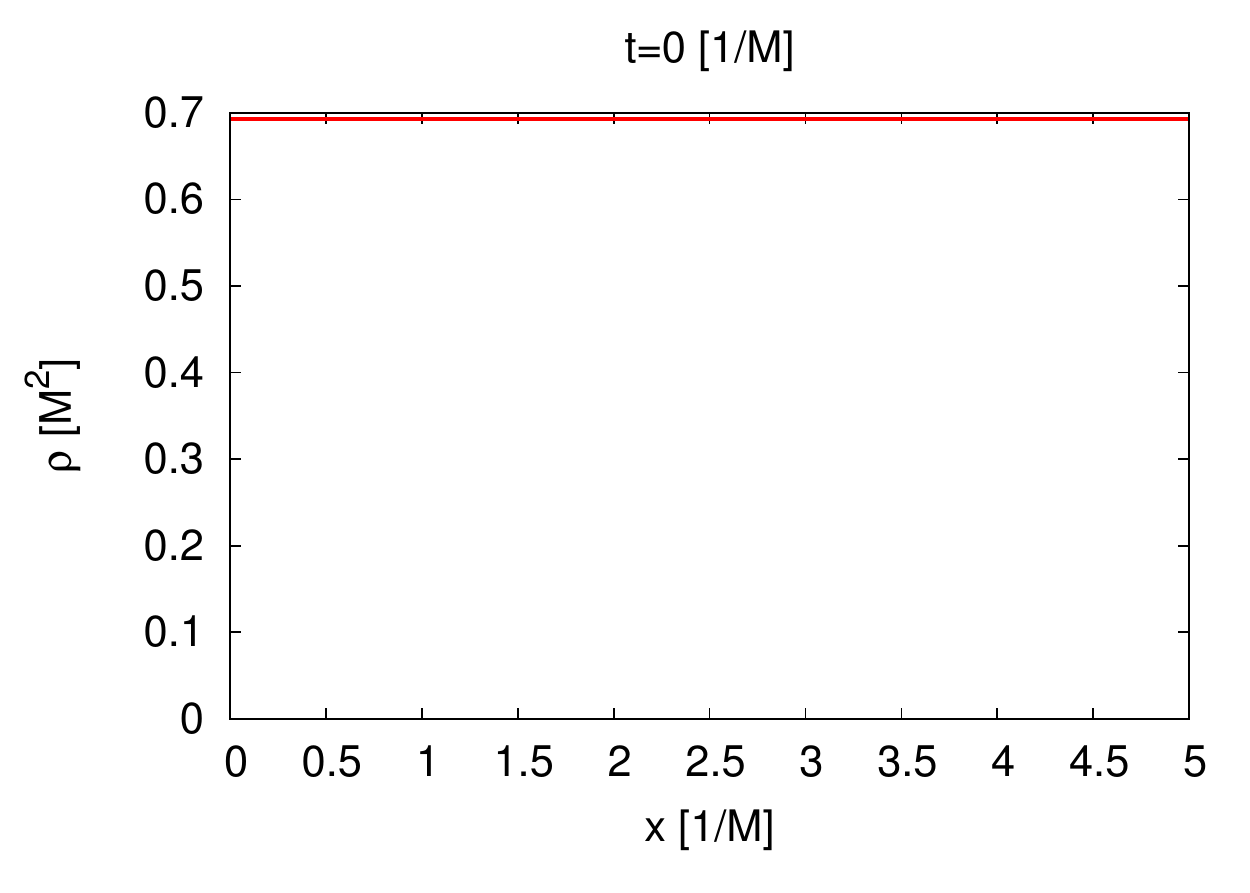}} &
\resizebox{80mm}{!}{\includegraphics{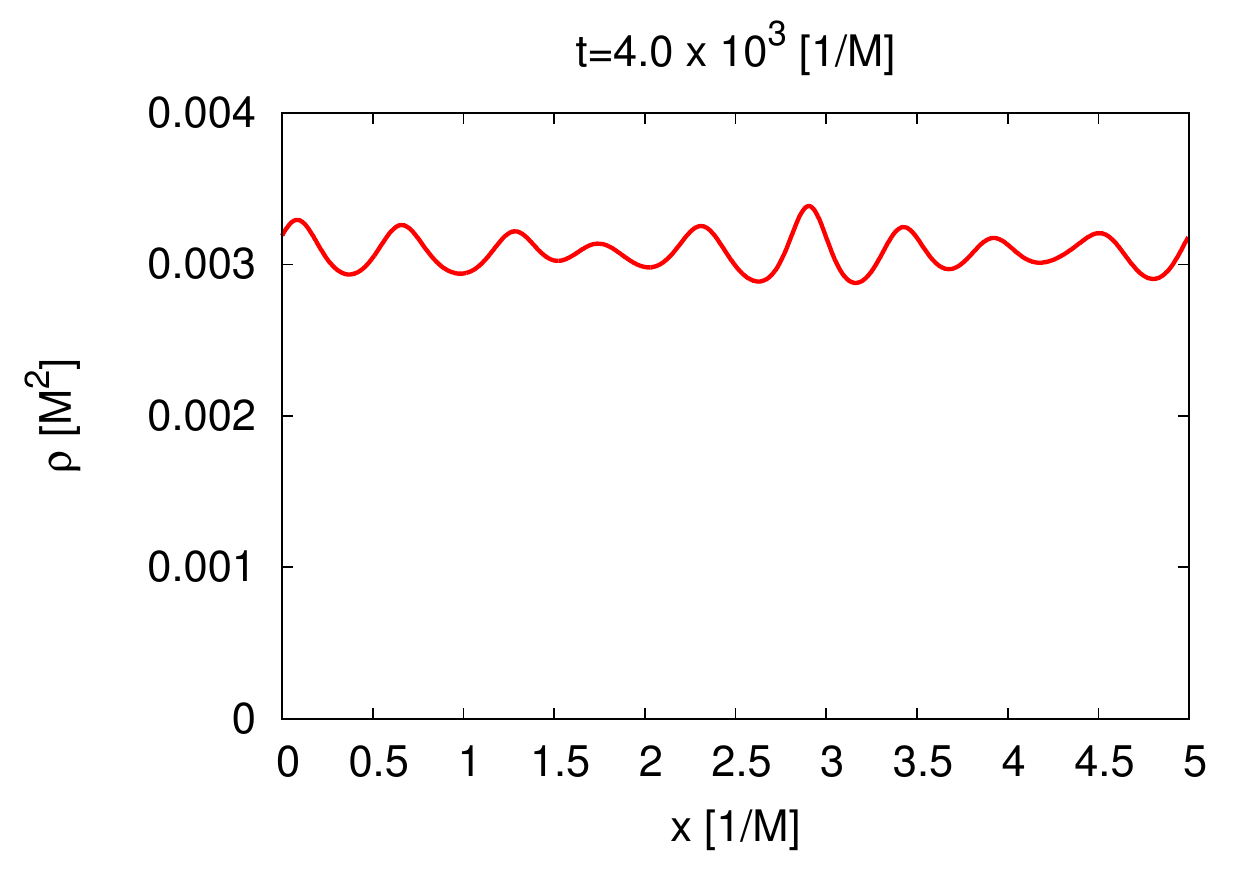}}\\
\resizebox{80mm}{!}{\includegraphics{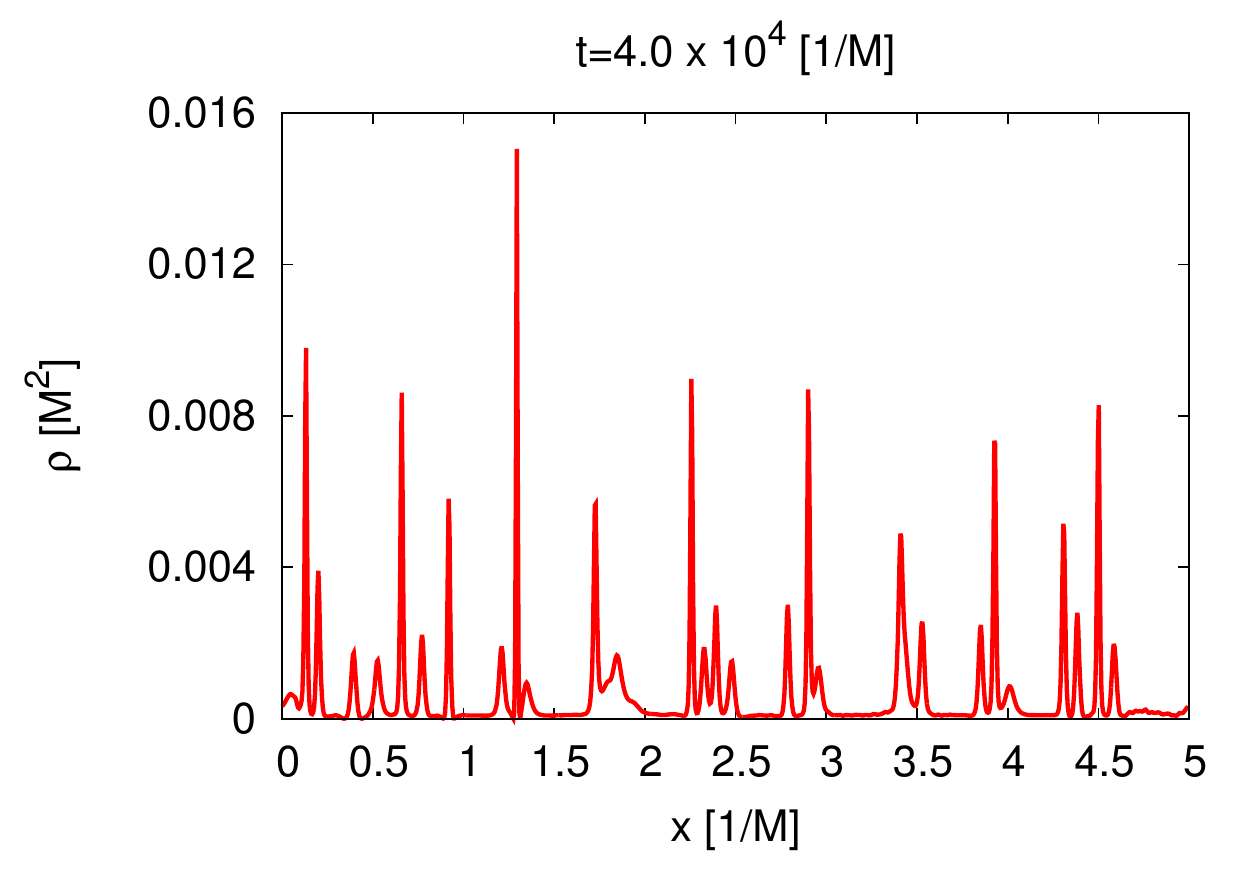}} &
\resizebox{80mm}{!}{\includegraphics{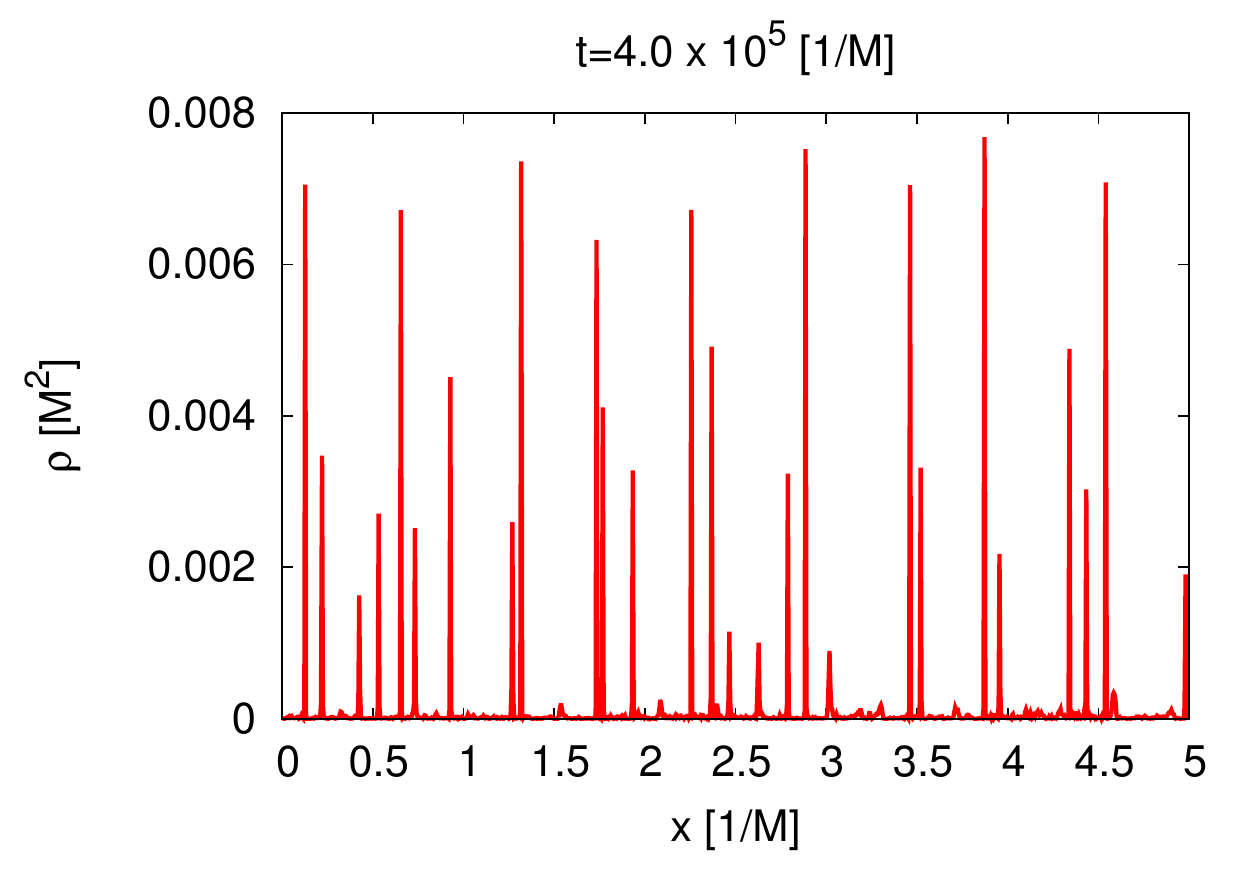}}\\
\end{tabular}
\caption{
Time evolution of the spatial distribution of the energy density $\rho$ 
from $t=0[1/M]$ to $t=4\times10^{5}[1/M]$. 
The initial amplitude is $\phi_0=1$. The lattice setting is as 
$L=5/M,~N=1024$. From the top left panel to the bottom right panel, the time 
evolves as $t=0,~4\times10^{3},~4\times10^{4},~4\times10^{5}[1/M]$ respectively. 
}
\label{FIG:1}
\end{center}
\end{figure}

\begin{figure}[htbp]
\begin{center}
\resizebox{150mm}{!}{\includegraphics{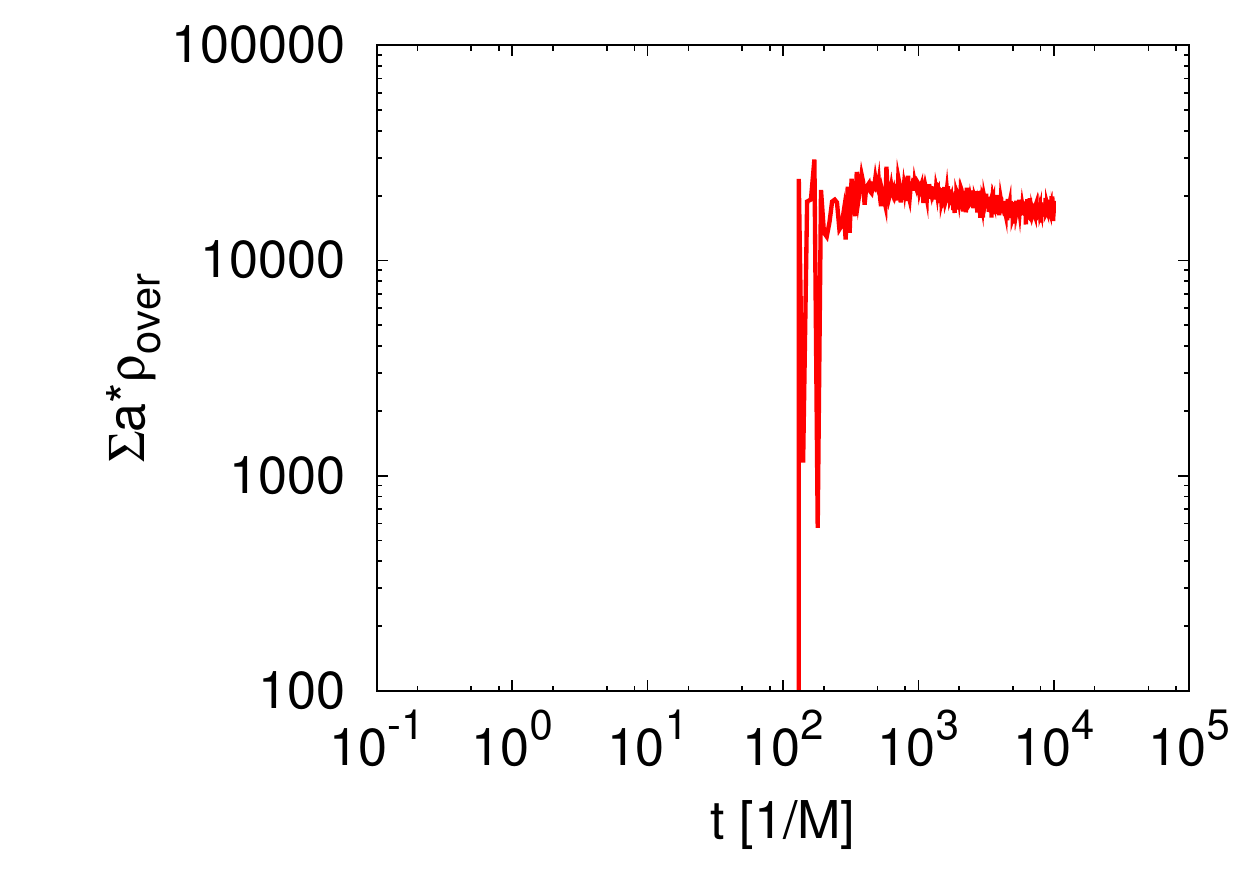}} 
\caption{
Time evolution of the sum of the over density where $\rho_{\rm over}$ in Lattice grid is defined as $\rho_{\rm over}>2\braket{\rho}$
from $t=10^{-1}[1/M]$ to $t=10^{5}[1/M]$ for $\phi_0(t_0)=10$.
}
\label{FIG:1_5}
\end{center}
\end{figure}

\begin{figure}[htbp]
\begin{center}
\begin{tabular}{c c}
\resizebox{80mm}{!}{\includegraphics{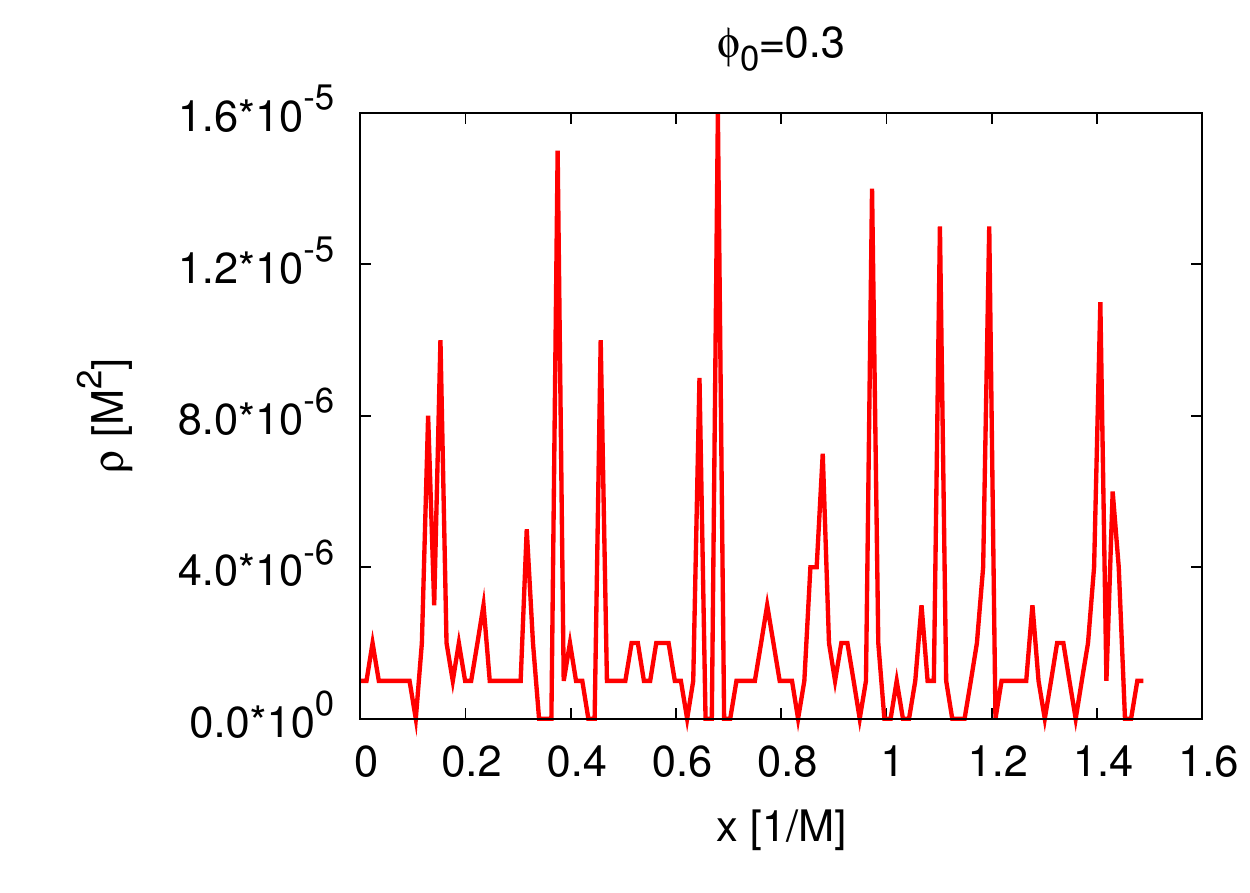}} &
\resizebox{80mm}{!}{\includegraphics{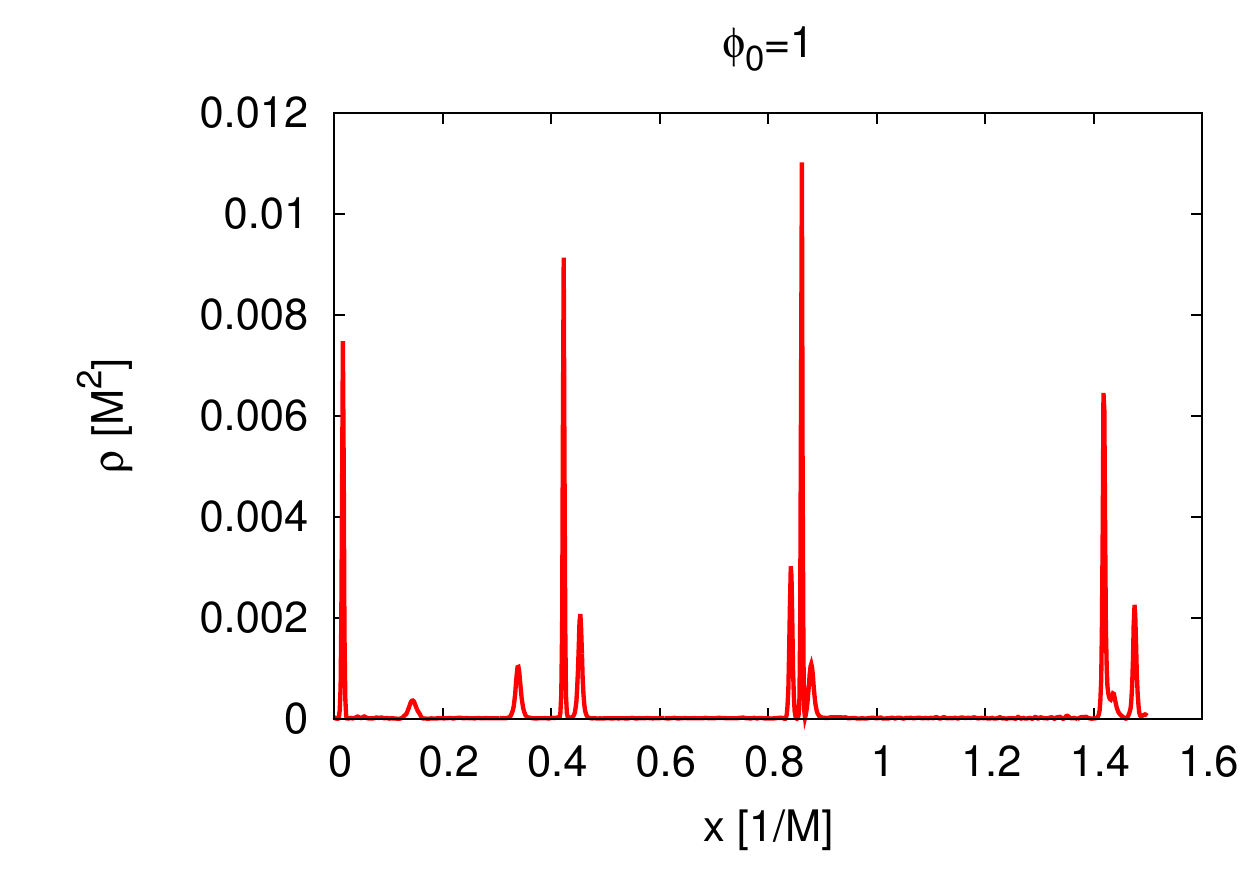}}\\
\resizebox{80mm}{!}{\includegraphics{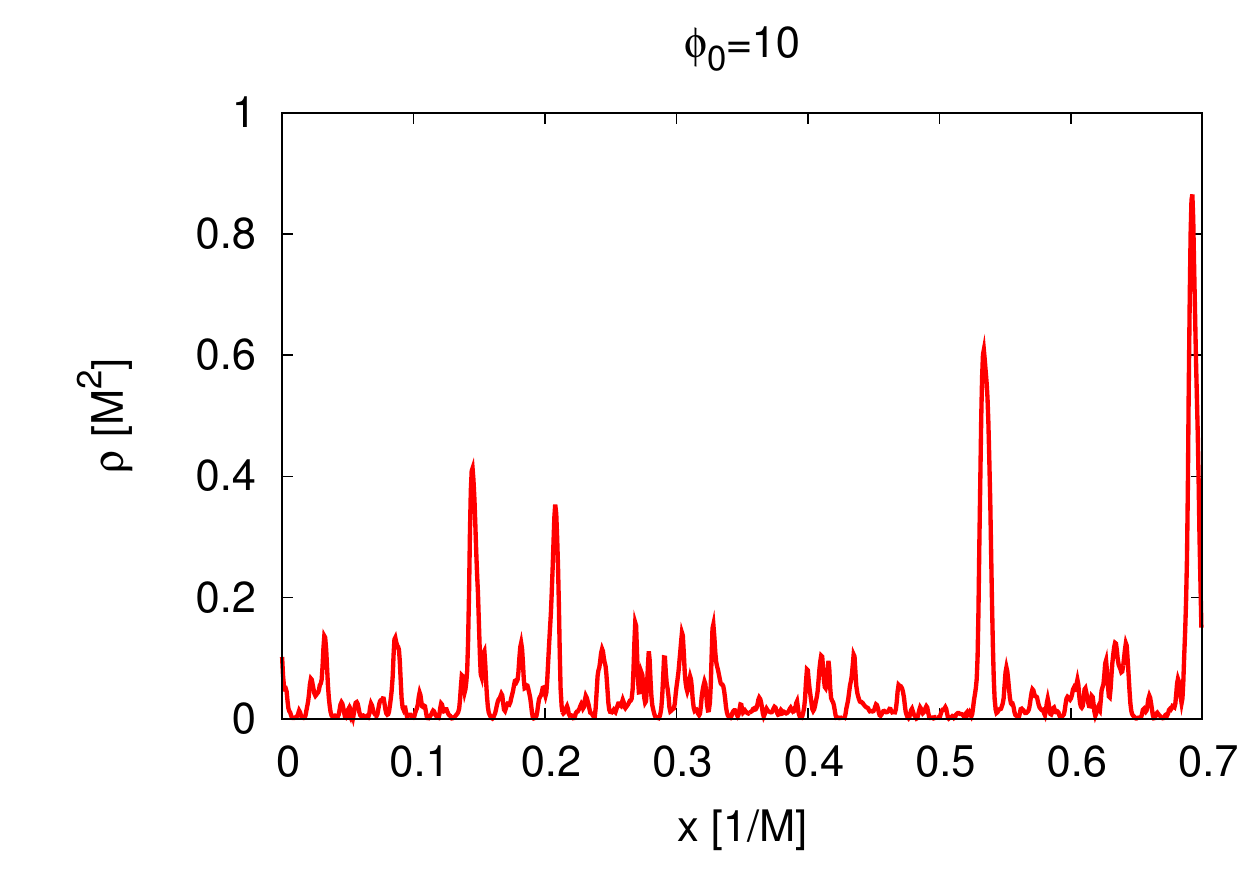}} &
\resizebox{80mm}{!}{\includegraphics{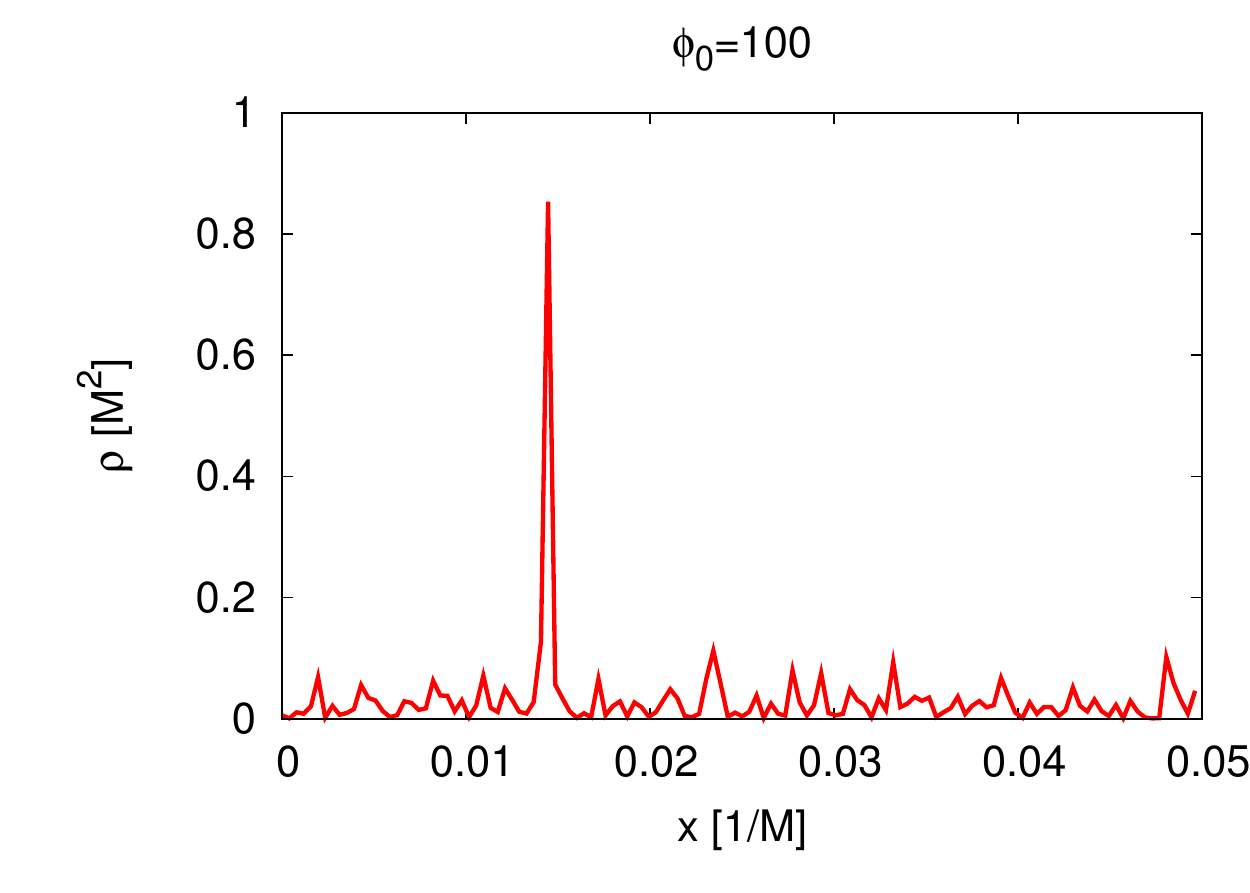}}\\
\end{tabular}
\caption{
Snapshot of the distribution of the energy density at each grid after the formation of the  I-ball
for each initial condition $\phi_0=0.3,1,10,100$.
}
\label{FIG:1_6}
\end{center}
\end{figure}

We present the result of the lattice simulation in Fig.\ref{FIG:1},  
which shows the time evolution of the spatial distribution of the energy density
$\rho=(1/2)[\dot{\phi}^2+(\nabla \phi)^2/a^2]+V$ from $t=0$ to $t=4\times10^5$
(in units of $1/M$). 
It is seen that the scalar field has fragmented into the spatially
localized and stable objects which are regarded as I-balls at $t=4\times10^{5}$. 
At first, the energy density $\rho$ decrease by 
the Hubble expansion and, at the same time, the fluctuations of the field 
are enhanced through parametric resonance. 
Then the field fluctuations fragment into I-balls after considerably lots of oscillations. 
During this process, the sum of the comoving energy density  which is larger than two times of the average energy density 
increases and reaches the nearly constant value as showed in Fig.~\ref{FIG:1_5} where $a$ is the  scale factor. 
Using this evolution of the sum of the comoving energy density,
we determined the time scale of the formation of the I-ball at the time when 
the sum of the over density reaches the almost constant value as like $\Delta t\simeq10^3[1/M]$.
After the formation of the I-ball, the spatial distribution of the energy density for each initial amplitude is showed in Fig.~\ref{FIG:1_6}.
After the formation of the I-ball, the energy density is conserved and then the amplitude of the field $\Phi$ is given as $\rho=\dot{\phi}^2/2+V=V(\Phi)$.
Using this relation, we obtain the amplitudes of the  I-balls and determine the typical amplitude of the I-ball as the maximum amplitude of them.
Using the time evolution of the energy as like Fig.~\ref{FIG:1_5} and the distribution of the energy density in Fig.~\ref{FIG:1_6},
we summarized the formation time $\Delta t$ and the typical amplitude $\Phi(0)$ in Table~\ref{table:1} where the values are order of magnitude estimations.

\begin{table}[htbp]
\scalebox{1.5}{
\begin{tabular}{|c|c|c|c|c|} \hline
$\phi_0$ & $0.3$ & $1$ & $10$ & $100$\\ \hline
$\Delta t[1/M]$ & $10^6$ & $10^5$ & $10^3$ & $10^4$ \\ \hline
$\Phi (0)$ & $10^{-3}$ & $10^{-1}$ & $1$ & $1$\\ \hline
\end{tabular}
}
\caption{
Typical value of I-ball formation. 
$\phi_0$ is the initial amplitude of the simulation. 
$\Delta t$ is the time to form I-ball. 
$\Phi (0)$ is the amplitude at the center of the I-ball.
}
\label{table:1}
\end{table}

With the logarithmic potential~(\ref{eq:simulation-1}), the inflection point is 
at $\phi=1$ and the deviation from the quadratic potential becomes 
significant beyond this point. 
Table~\ref{table:1}, we can see that even if the scalar 
field starts to roll down the potential with the initial amplitude 
$\phi_0 \gg 1$, it fragment into I-balls. 
In addition, we see that the formation time of I-balls becomes shorter 
as the initial amplitude is larger for $\phi_0<10$, but becomes longer for $\phi_0>10$.
This amplitude dependence of the formation time is related with the growth rate of the fluctuations by the 
parametric resonance. 
In logarithmic potential, the amplification of the fluctuations per one oscillation becomes larger as the amplitude 
is larger, but the period of the one oscillation becomes larger for the larger amplitude than $1$. 
As a consequence, the growth rate of fluctuation in unit time of $1/M$ becomes larger as the amplitude is larger for $\phi_0 \lesssim 10$, but
becomes smaller for $\phi_0 \gtrsim 10$, which leads to the amplitude dependence of the formation time.
In the case for $\phi_0\gg1$, although the fluctuations of the field are enhanced for at 
the large amplitude of the homogeneous mode and the fragmentation starts, the formation of 
the I-ball (stable configuration of the filed) is completed when the amplitude 
drops to ${\cal O}(1)$ due to the Hubble expansion. 
Furthermore, we have executed the simulation with no expansion. 
In that case, we have not confirmed the formation of the I-ball for the large 
amplitude of the field $\phi>10$. 
In all previously known cases, I-balls are formed with the 
quasi-quadratic potential~\cite{Gleiser:1993pt}. 
For the case of the logarithmic potential,  the I-balls are also formed 
when the scalar field oscillates at the region 
where the potential is approximately given by the quadratic form, which 
results in $\Phi (0)\le 1$ as seen in Table~\ref{table:1}. 
Thus, the oscillation amplitude of the I-ball is limited above.
The fact that I-balls are formed when the quadratic potential is dominant
is consistent with the idea that the adiabatic invariance is crucial in
the I-ball formation.  


\section{Analytical estimate}
\label{sec:Analytical estimate}


Although it is not fully understood what makes the I-ball stable,
Ref.~\cite{Kasuya:2002zs} pointed out that the adiabatic invariance plays 
the crucial role for its formation and the result of the simulation 
in the previous section is consistent with this. 
In the classical mechanics, if the system undergoes a periodic 
motion with some external force which is adiabatic enough not to disturb 
the periodic motion, there exists the adiabatic 
invariant, namely, the area of the track in the phase space is 
conserved~\cite{Landau:1976}. 
This can be extended to the classical fields~\cite{Kasuya:2002zs} 
and the adiabatic invariant is defined as
\begin{equation}\label{eq:ana-0}
   I\equiv\frac{1}{2}\frac{T}{2\pi}\int dx~\bar{\dot{\phi}}^2,
\end{equation}
where the over line represents the average over the period of the motion and 
$T$ is the period of the oscillation.
In this section, we analytically estimate the configuration of the
I-ball under the assumption that the adiabatic invariant~(\ref{eq:ana-0}) is conserved 
during the I-ball formation. 
We also investigate the instability mode of the fluctuations 
and consider the relation to the radius of the I-ball.

\subsection{PROFILE OF I-BALL}

In the previous studies, the configuration of the I-ball was estimated by expanding the amplitude of the field
by small parameter $\epsilon$~\cite{Fodor:2009kf,Farhi:2007wj,Dashen:1975}. 
Although it describes the I-ball profile very well,  the physical 
reason for the existence of the I-ball is not clear. 
Furthermore, the $\epsilon$ expansion cannot be applied to some potentials such as
$V \sim \phi^{2-K}$ with $1 \gg K>0$ which has an I-ball solution~\cite{Kasuya:2002zs}. 
Thus, in this paper, instead of using $\epsilon$ expansion, we estimate the profile assuming the 
conservation of the adiabatic invariant for the formation of I-ball.
This method gives the same profile of the I-ball obtained from the small expansion method.

\begin{figure}[htbp]
\begin{center}
\resizebox{120mm}{!}{\includegraphics{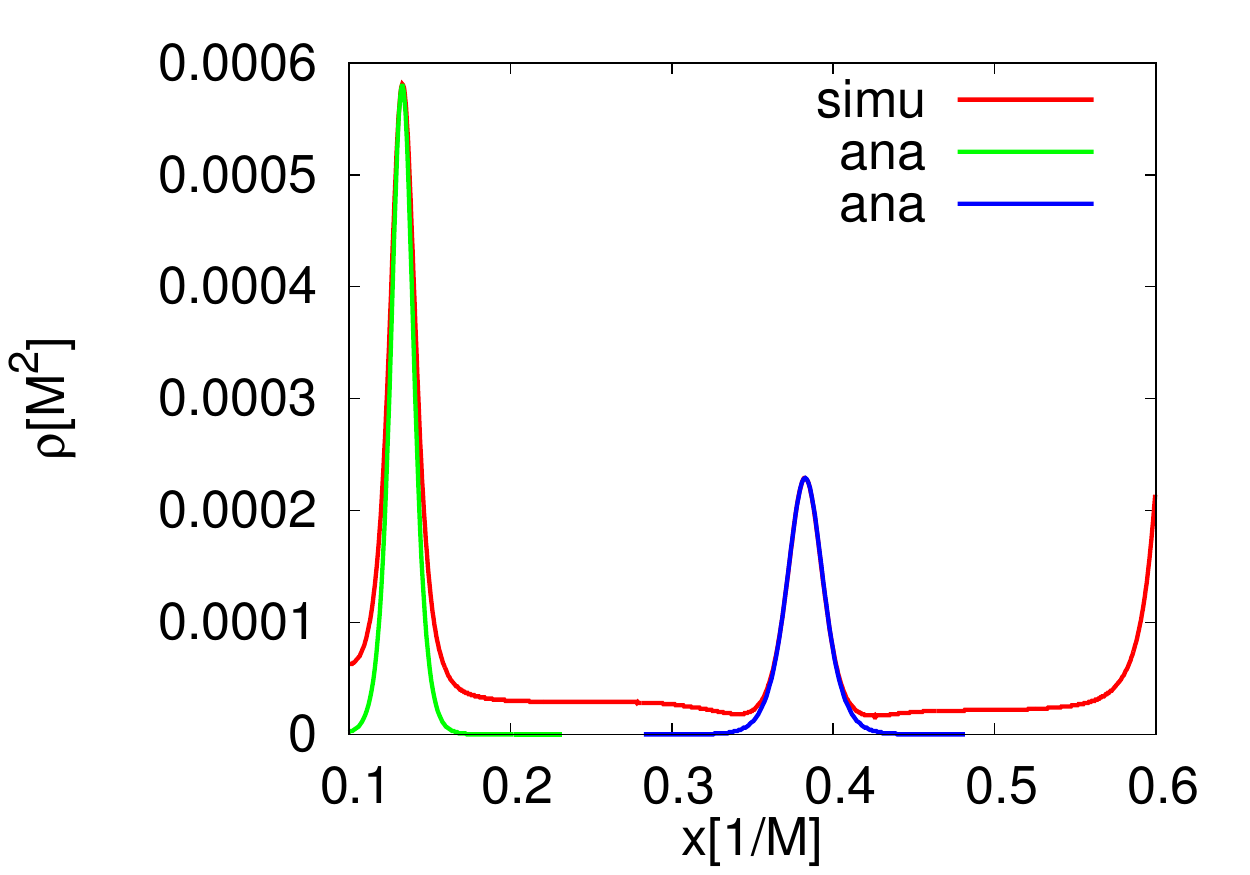}}
\caption{Comparison of I-ball configuration of the analytical estimation
(\ref{eq:Phi_profile})(blue and green line)
with the result of the simulation (red line). 
}
\label{FIG:2}
\end{center}
\end{figure}

We presume that the configuration of the I-ball takes 
the lowest energy with fixed $I$ when it is formed. 
In this situation, using the method of the Lagrange multipliers, 
we can derive the field configuration by minimizing $E_{\omega}$ defined as 
\begin{equation}\label{eq:ana-1}
\begin{split}
   E_{\omega}&\equiv
   \overline{E}
   +\tilde{\omega}\left(I-\frac{1}{2}\frac{T}{2\pi}\int dx~\overline{\dot{\phi}^2}\right)
   \\[0.6em]
   &=\int dx\left[
   \left(1-\frac{T}{2\pi}\tilde{\omega}\right)\frac{1}{2}\overline{\dot{\phi}^2}
   +\frac{1}{2}\overline{(\nabla\phi)^2}
   +\overline{V(\phi)}
   \right].
\end{split}
\end{equation}
The quantities averaged over the period are given by
\begin{equation}\label{eq:ana-2}
\begin{split}
   &\overline{\phi^2}\simeq\frac{1}{2}\Phi^2,\\
   &
   \overline{\dot{\phi}^2}\simeq\frac{1}{2}\times(\sqrt{2}M)^2\Phi^2
   ,\\
   &
   \overline{V(\phi)}=\overline{M^2\log\left[1+\phi^2\right]}
   \simeq M^2\left(\frac{1}{2}\Phi^2-\frac{1}{2}\frac{3}{8}\Phi^4\right)
\end{split}
\end{equation}
where $\Phi$ is the amplitude of the oscillation and 
we approximate the frequency of the oscillation 
as $2\pi/T\simeq\sqrt{2}M$. 
Variation of $E_{\omega}$~(\ref{eq:ana-1}) with respect to $\Phi$ leads to 
\begin{equation}\label{eq:ana-3}
   \frac{d^2}{dr^2}\Phi +(\sqrt{2}\omega-2M)M\Phi+\frac{3}{2}M^2\Phi^3=0,
\end{equation}
where $\omega\equiv \tilde{\omega}-\sqrt{2}M$. 
For eq.~(\ref{eq:ana-3}), 
we can obtain the analytical solution as 
\begin{equation}\label{eq:Phi_profile}
   \Phi(r)=\Phi(0){\rm sech}\left(\frac{\sqrt{3}\Phi(0)}{2}Mr\right),
\end{equation}
where $\Phi(0)$ is the amplitude at the center of the I-ball. 
This estimated profile~(\ref{eq:Phi_profile}) accords with that by $\epsilon$ expansion (Appendix~
\ref{sec:e expansion}). 
From the above configuration (\ref{eq:Phi_profile}), 
we can estimate the size of the I-ball as 
\begin{equation}\label{eq:iball_radius}
   R=\frac{1.6}{\Phi(0)}\frac{1}{M},
\end{equation}
where $R$ is the distance from the center at which $\Phi(R) = \Phi(0)/2$. 
Fig.~\ref{FIG:2} shows the comparison of I-ball 
configuration of the analytical estimation~(\ref{eq:Phi_profile}) 
with the result of the simulation. 
From this comparison, we can see that the I-ball configuration obtained by 
minimizing the energy $E_{\omega}$ under the existence of adiabatic invariant $I$ 
describes the result of the simulation quite well. 

\subsection{INSTABILITY}
In the process of the I-ball formation, at first the fluctuations are enhanced by the parametric resonance 
due to the oscillation of the field, and then the fluctuations fragment into the I-ball. 
In this subsection, we derive the instability mode of the log potential~(\ref{eq:simulation-1}) and  
compare it with the radius of the 
I-ball~(\ref{eq:iball_radius}). 

In the Fourier space, the equation of motion for the fluctuation 
mode $\delta\phi_{\mathbf{k}}$ is given by 
\begin{equation}\label{eq:fluctuation}
   \delta\ddot{\phi}_{\mathbf{k}}
   +\left[k^2+2M^2\frac{1-\phi^2_0(t)}{(1+\phi_0^2(t))^2}\right]\delta\phi_{\mathbf{k}}=0,
\end{equation}
where we decomposed the field into the background and the fluctuation as 
$\phi(t,x)=\phi_0(t)+\delta\phi(t,x)$ and ignored the cosmic expansion. 

For small amplitude of the background fied, we can approximate the eq.~(\ref{eq:fluctuation}) as
\begin{equation}\label{eq:fluctuation_mode}
\delta\ddot{\phi}_{\mathbf{k}}
   +\left[k^2+2M^2(1-\phi^2_0(t))\right]\delta\phi_{\mathbf{k}}=0
\end{equation}
Since the background $\phi_0$ oscillates 
with frequency~$\sqrt{2}M$ as $\phi_0(t)=\Phi\cos(\sqrt{2}Mt)$, 
the eq.~(\ref{eq:fluctuation_mode}) is written as 
\begin{equation}\label{eq:Mathieu}
   \frac{d^2}{d\tau^2}\delta\phi_{\mathbf{k}}+
   \left[
      \frac{k^2+(2-\Phi^2)}{2}M^2-\frac{\Phi^2}{2}M^2\cos(2M\tau)
   \right]\delta\phi_{\mathbf{k}}=0,
\end{equation}
where $\tau$ is defined as $\tau\equiv \sqrt{2}t$. 
This is the Mathieu equation~\cite{Gradshteyn:1995} 
which has instability mode at $(k^2/M^2+2-\Phi^2)/2\simeq1$ 
(e.g. see the stability/instability chart in~\cite{Kofman:1994rk}). 
Thus, the instability occurs at 
\begin{equation}
   \frac{1}{k}\sim\frac{1}{\Phi}\frac{1}{M}.
\end{equation}
This scale is in accord with the result of the estimation for radius of I-ball~
(\ref{eq:iball_radius}).

\begin{figure}[htbp]
\begin{center}
\begin{tabular}{c c}
\resizebox{80mm}{!}{\includegraphics{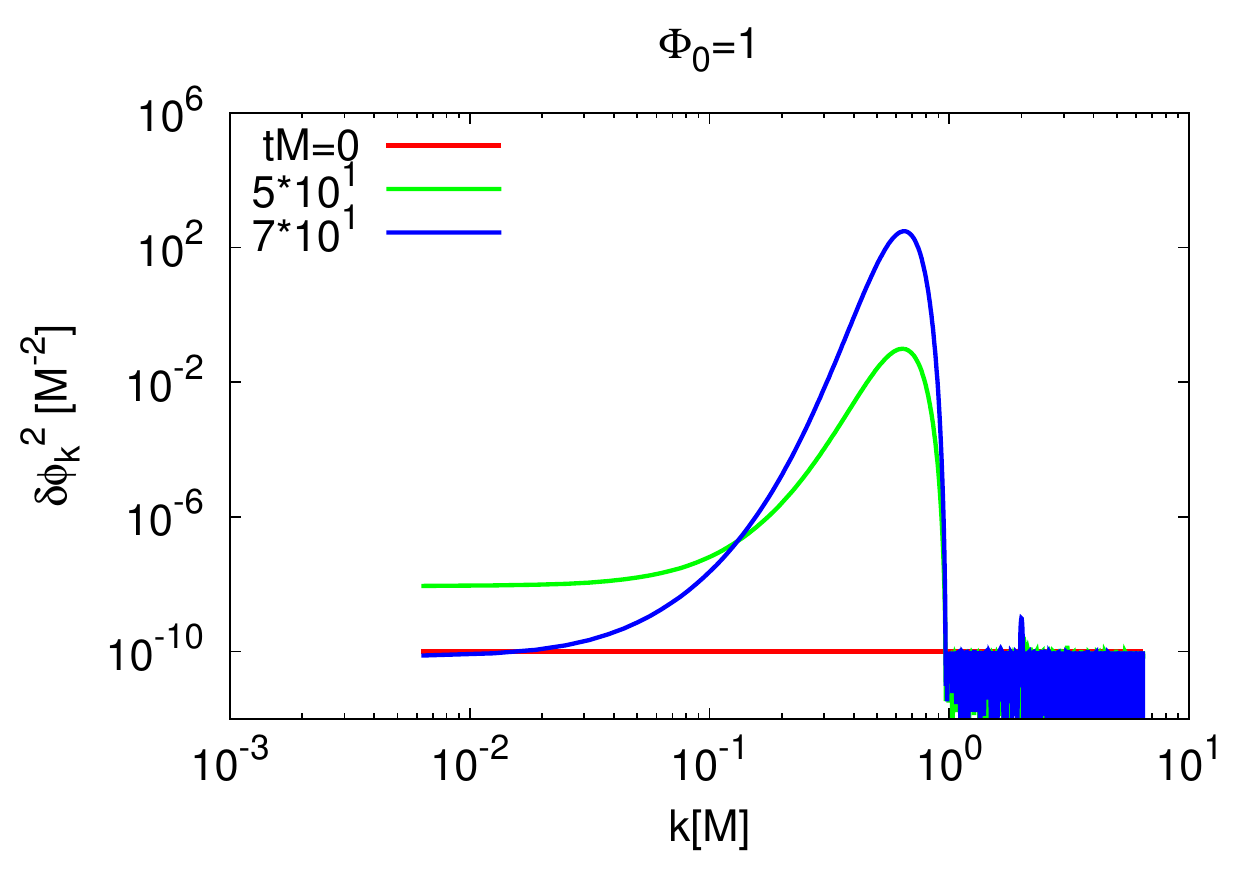}} &
\resizebox{80mm}{!}{\includegraphics{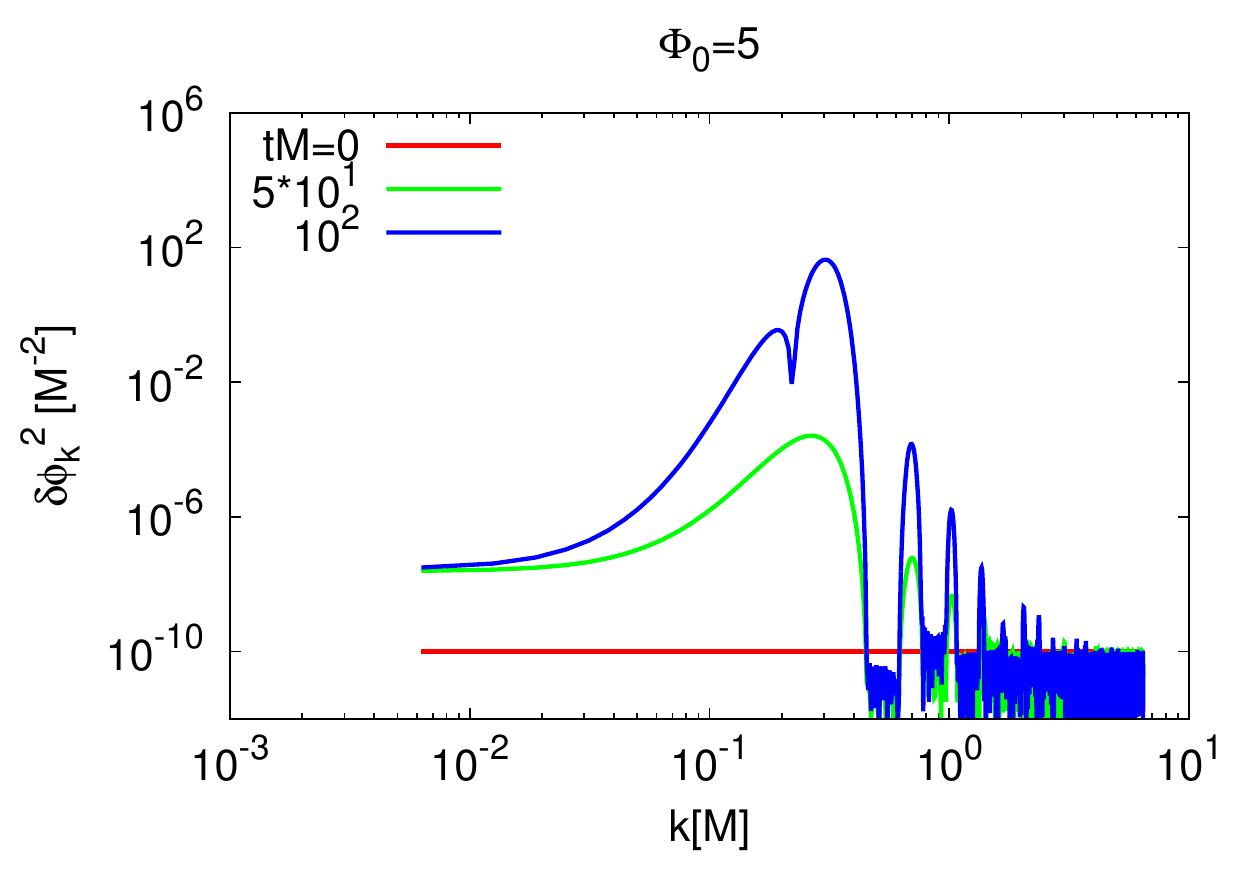}}\\
\resizebox{80mm}{!}{\includegraphics{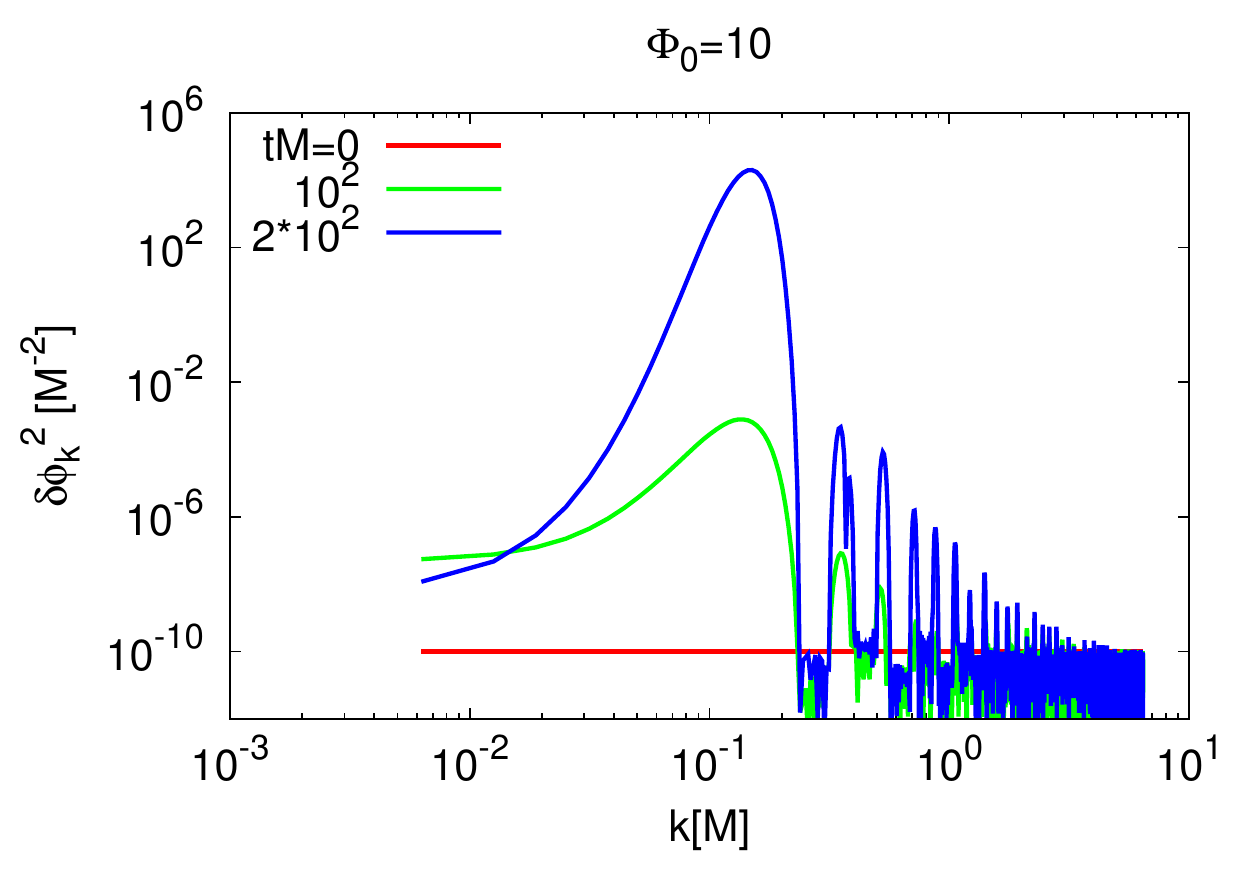}} &
\resizebox{80mm}{!}{\includegraphics{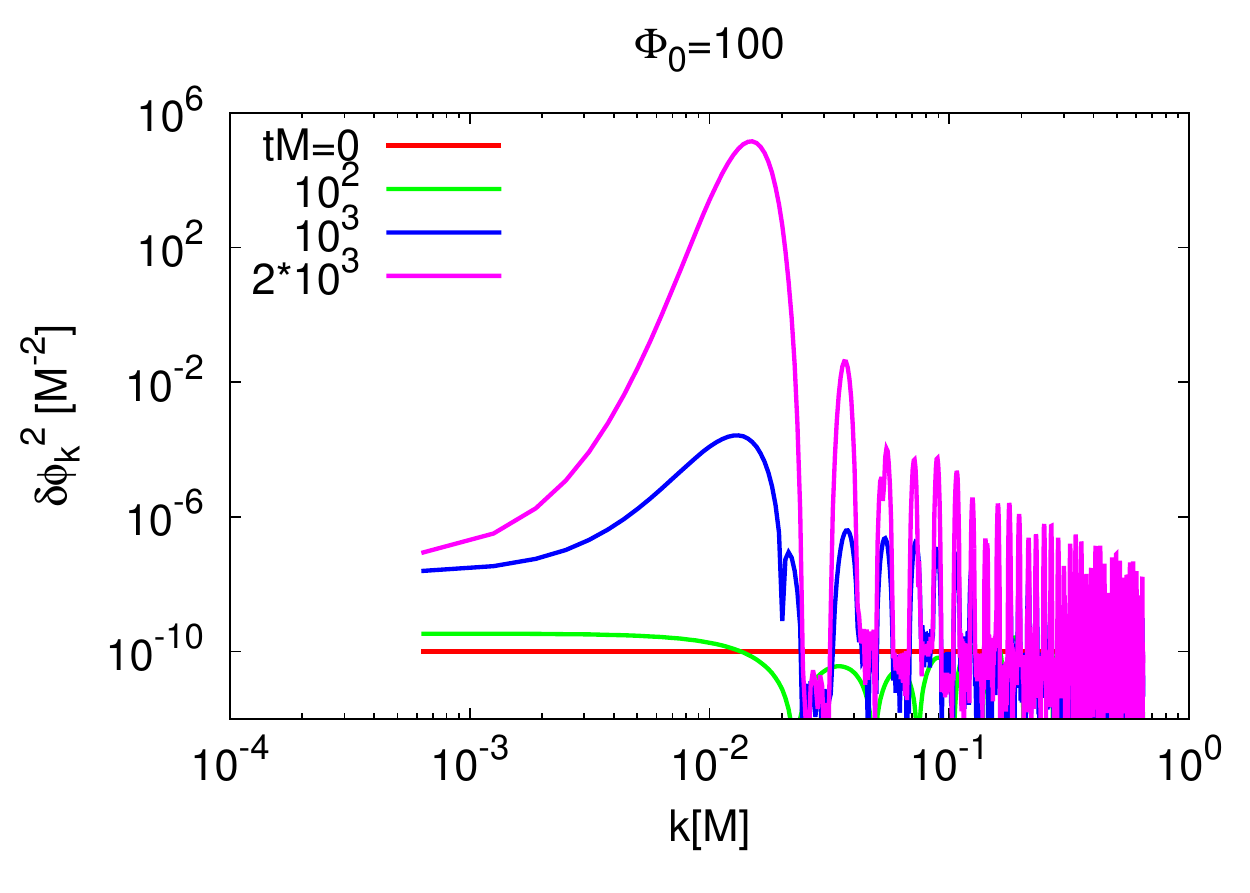}} 
\end{tabular}
\caption{
Evolution of instabilities of the logarithmic potential for the amplitude of the background
$\phi_0=1,5,10,100$.
The vertical axis is the amplitude of the fluctuation and the horizontal axis is the corresponding momentum.
}
\label{FIG:3}
\end{center}
\end{figure}

\begin{figure}[htbp]
\begin{center}
\begin{tabular}{c c}
\resizebox{80mm}{!}{\includegraphics{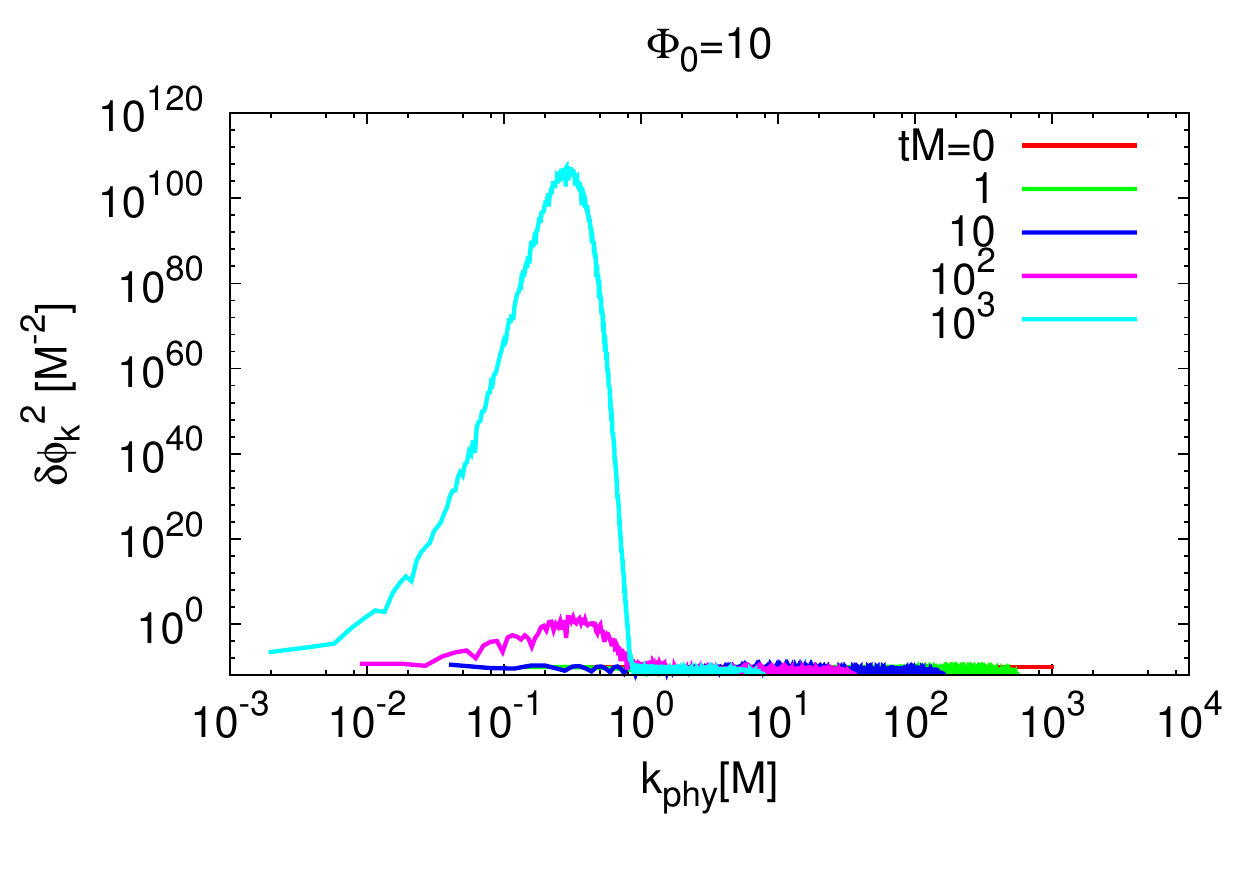}} &
\resizebox{80mm}{!}{\includegraphics{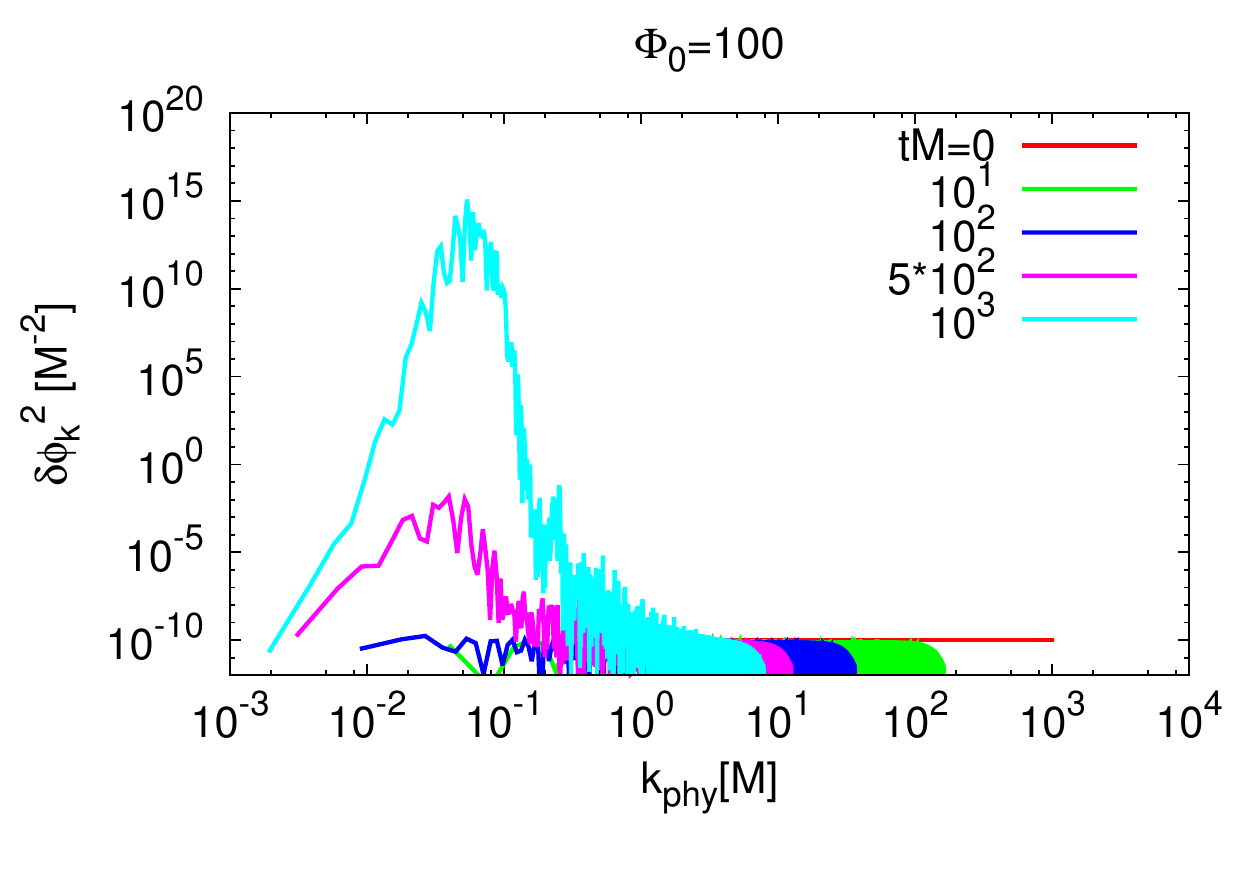}}\\
\end{tabular}
\caption{
Evolution of instabilities of the logarithmic potential under the Hubble expansion for large initial amplitude
$\phi_0=10,100$.
The vertical axis is the amplitude of the fluctuation and the horizontal axis is the corresponding physical momentum.
}
\label{FIG:3.5}
\end{center}
\end{figure}

For the larger amplitude than 1, we can not perturbatively expand the potential as in 
(\ref{eq:fluctuation_mode}). 
In this case, we have studied the instability numerically solving eq.~(\ref{eq:fluctuation}) and results are showed in Fig.~\ref{FIG:3}.
From Fig.~\ref{FIG:3}, we can see that for the larger amplitude than 1, the instability occurs at several modes and that the most growing mode is 
tipycally $k\sim{\cal O}(0.01)-{\cal O}(0.1)[1/M]$.
These multi instability modes have the possibility to obstacle the formation of the I-ball at the much large amplitude $\Phi>>1$, however, 
under the Hubble expansion, these multi instabilities but one mode damp
as showed in Fig.~\ref{FIG:3.5} 
where we have numerically solved the e.o.m. for background and fluctuation as
\begin{equation}
\ddot{\phi}_0+H\dot{\phi}_0+\frac{\partial V}{\partial \phi}=0
\end{equation}
and 
\begin{equation}
\delta\ddot{\phi}_k+H\delta\dot{\phi}_k+\left[\frac{k^2}{a^2}+\frac{\partial^2 V}{\partial\phi^2}(\phi_0)\right]\delta\phi_k=0.
\end{equation}
Correspondence of the instability mode with the size of the formed I-ball (Table~\ref{table:1} and eq.~(\ref{eq:iball_radius})) 
suggests that the enhanced fluctuations by the parametric resonance seeds for the formation of the I-ball.

So far we see the instability in eq.~(\ref{eq:fluctuation}) which is linear in 
$\delta\phi$.
When the fluctuations increase, the non-linear effect becomes important.
The accordance between the instability mode and the radius of the I-ball
suggests  that I-ball is formed with the balance between the pressure 
coming from the gradient term $(\nabla\phi)^2$ and enhanced fluctuations 
of the fields by the parametric resonance. 


\section{CONCLUSION}
\label{sec:CONCLUSION}


In this paper, we have confirmed that a coherently oscillating field with 
logarithmic potential fragments into I-balls using lattice simulation. 
However, the I-balls are formed after the scalar potential becomes
dominated by the quadratic term.
As a consequence, the amplitude of I-ball is limited above. 
This result suggests that the adiabatic invariance 
plays an import role in the I-ball formation. 
In fact, we have estimated the I-ball configuration under the assumption 
that the adiabatic invariant is conserved, and 
the estimated configuration is well fitted 
to the result of the simulation. 
This logarithmic potential appears in the various situations in the early Universe 
and hence the I-ball formation would affect the cosmological scenario, 
which will be studied in a forthcoming paper~\cite{forthcoming}.

\section*{Acknowledgments}
This work is supported by Grant-in-Aid for Scientific research
from the Ministry of Education, Science, Sports and Culture
(MEXT), Japan, No. 25400248 (M.K.), No. 21111006 (M.K.)
and also by World Premier International Research Center Initiative
(WPI Initiative), MEXT, Japan.
\appendix
\section{$\epsilon$ expansion}
\label{sec:e expansion}
In this section, we estimate the profile of the I-ball using $\epsilon-$expansion
\cite{Fodor:2009kf,Farhi:2007wj,Dashen:1975} and compare the 
result with that by I conservation. 
The equation of motion for the field $\phi$ is given as
\begin{equation}
\label{eq:A-1}
\frac{d^2}{dt^2}\phi-\frac{d^2}{dx^2}\phi+\frac{\partial V}{\partial \phi}=0. 
\end{equation}
We assume that the amplitude of the I-ball is small and the deviation of the frequency from that 
of the quadratic potential is small 
when I-ball is formed. 
Under this assumptions, we expand 
the field by small parameter $\epsilon$ and change the variable as 
\begin{eqnarray}
\label{eq:A-2}
\phi=\epsilon \phi_1 + \epsilon^2 \phi_2 +\epsilon^3 \phi_3 +{\cal O}(\epsilon^4),\\
\label{eq:A-3}
\tau\equiv\sqrt{2}Mt\sqrt{1-\epsilon^2},~\chi\equiv\sqrt{2}Mx\epsilon.
\end{eqnarray}
Substituting~(\ref{eq:A-2})~(\ref{eq:A-3}) into~(\ref{eq:A-1}), we can obtain e.o.m. for $\phi_1~\phi_3$, 
as 
\begin{eqnarray}
\label{eq:A-4}
\phi_{1\tau\tau}+\phi_1&=&0,\\
\label{eq:A-5}
\phi_{3\tau\tau}+\phi_3&=&\phi_{1\tau\tau}+\phi_{1\chi\chi}+\phi_1^3.
\end{eqnarray}
Solving the eq.~(\ref{eq:A-4}), eq.~(\ref{eq:A-5}) reduces as 
\begin{equation}
\label{eq:A-6}
\phi_{3\tau\tau}+\phi_3=\left[f_{\chi\chi}-f+\frac{3}{4}f^3\right]\sin(\tau)-\frac{1}{4}f^3\sin(3\tau).
\end{equation}
where $\phi_1(\chi,\tau)=f(\chi)\sin(\tau)$. 
If the I-ball is stable, the first term of the right hand side of eq.~(\ref{eq:A-6}) should be $0$. 
\begin{equation}
\label{eq:A-7}
f_{\chi\chi}-f+\frac{3}{4}f^3=0.
\end{equation}
From the above eq.~(\ref{eq:A-7}), we can get the profile of the I-ball as 
\begin{equation}
\label{eq:A-8}
\Phi(r) = f(\sqrt{2}M\epsilon r) = \Phi(0){\rm sech}(\frac{\sqrt{3}}{2}\Phi(0)Mr). 
\end{equation}
This profile~(\ref{eq:A-8}) accords with the profile~(\ref{eq:Phi_profile}) obtained from the I-conservation 
assumption. 


\begin{thebibliography}{99}
\bibitem{Kibble:1976sj} 
  T.~W.~B.~Kibble,
  J.\ Phys.\ A {\bf 9}, 1387 (1976).

\bibitem{Coleman:1985ki} 
  S.~R.~Coleman,
  Nucl.\ Phys.\ B {\bf 262}, 263 (1985)
  [Erratum-ibid.\ B {\bf 269}, 744 (1986)].
  
\bibitem{Zeldovich:1974uw} 
  Y.~.B.~Zeldovich, I.~Y.~.Kobzarev and L.~B.~Okun,
  Zh.\ Eksp.\ Teor.\ Fiz.\  {\bf 67}, 3 (1974)
  [Sov.\ Phys.\ JETP {\bf 40}, 1 (1974)].

\bibitem{Kusenko:1997si} 
  A.~Kusenko and M.~E.~Shaposhnikov,
  Phys.\ Lett.\ B {\bf 418}, 46 (1998)
  [hep-ph/9709492].

\bibitem{Enqvist:1997si} 
  K.~Enqvist and J.~McDonald,
  Phys.\ Lett.\ B {\bf 425}, 309 (1998)
  [hep-ph/9711514].

\bibitem{Enqvist:1998en} 
  K.~Enqvist and J.~McDonald,
  Nucl.\ Phys.\ B {\bf 538}, 321 (1999)
  [hep-ph/9803380].
  
\bibitem{Kasuya:1999wu} 
  S.~Kasuya and M.~Kawasaki,
  Phys.\ Rev.\ D {\bf 61}, 041301 (2000)
  [hep-ph/9909509].

\bibitem{Kasuya:2000wx} 
  S.~Kasuya and M.~Kawasaki,
  Phys.\ Rev.\ D {\bf 62}, 023512 (2000)
  [hep-ph/0002285].

\bibitem{Kasuya:2001hg} 
  S.~Kasuya and M.~Kawasaki,
  Phys.\ Rev.\ D {\bf 64}, 123515 (2001)
  [hep-ph/0106119].
        
\bibitem{Bogolyubsky:1976nx} 
  I.~L.~Bogolyubsky and V.~G.~Makhankov,
  JETP Lett.\  {\bf 24}, 12 (1976).
  
\bibitem{Kasuya:2002zs} 
  S.~Kasuya, M.~Kawasaki and F.~Takahashi,
  Phys.\ Lett.\ B {\bf 559}, 99 (2003)
  [hep-ph/0209358].
  
\bibitem{Amin:2013ika} 
  M.~A.~Amin,
  Phys.\ Rev.\ D {\bf 87}, 123505 (2013)
  [arXiv:1303.1102 [astro-ph.CO]].

\bibitem{Ade:2014xna} 
  P.~A.~R.~Ade {\it et al.}  [BICEP2 Collaboration],
  arXiv:1403.3985 [astro-ph.CO].

\bibitem{Ellis:2014rxa} 
  J.~Ellis, M.~A.~G.~Garcia, D.~V.~Nanopoulos and K.~A.~Olive,
  arXiv:1403.7518 [hep-ph].
  
  
  
  
  


  
\bibitem{Amin:2010xe} 
  M.~A.~Amin,
  arXiv:1006.3075 [astro-ph.CO].
 
\bibitem{Amin:2011hj} 
  M.~A.~Amin, R.~Easther, H.~Finkel, R.~Flauger and M.~P.~Hertzberg,
  Phys.\ Rev.\ Lett.\  {\bf 108}, 241302 (2012)
  [arXiv:1106.3335 [astro-ph.CO]].
  
\bibitem{McDonald:2001iv} 
  J.~McDonald,
  Phys.\ Rev.\ D {\bf 66}, 043525 (2002)
  [hep-ph/0105235].
  
\bibitem{Hertzberg:2010yz} 
  M.~P.~Hertzberg,
  Phys.\ Rev.\ D {\bf 82}, 045022 (2010)
  [arXiv:1003.3459 [hep-th]].
  
\bibitem{Zhou:2013tsa} 
  S.~-Y.~Zhou, E.~J.~Copeland, R.~Easther, H.~Finkel, Z.~-G.~Mou and P.~M.~Saffin,
  arXiv:1304.6094 [astro-ph.CO].


\bibitem{de Gouvea:1997tn} 
  A.~de Gouvea, T.~Moroi and H.~Murayama,
  Phys.\ Rev.\ D {\bf 56}, 1281 (1997)
  [hep-ph/9701244].
    
\bibitem{Mukaida:2012qn} 
  K.~Mukaida and K.~Nakayama,
  JCAP {\bf 1301}, 017 (2013)
  [JCAP {\bf 1301}, 017 (2013)]
  [arXiv:1208.3399 [hep-ph]].

\bibitem{Enqvist:2001zp} 
  K.~Enqvist and M.~S.~Sloth,  
  Nucl.\ Phys.\ B {\bf 626}, 395 (2002)
  [hep-ph/0109214].
  
\bibitem{Lyth:2001nq} 
  D.~H.~Lyth and D.~Wands,
  Phys.\ Lett.\ B {\bf 524}, 5 (2002)
  [hep-ph/0110002].
  
\bibitem{Moroi:2001ct} 
  T.~Moroi and T.~Takahashi,
  Phys.\ Lett.\ B {\bf 522}, 215 (2001)
  [Erratum-ibid.\ B {\bf 539}, 303 (2002)]
  [hep-ph/0110096].
  
\bibitem{Murayama:1992ua} 
  H.~Murayama, H.~Suzuki, T.~Yanagida and J.~'i.~Yokoyama,  
  Phys.\ Rev.\ Lett.\  {\bf 70}, 1912 (1993).

\bibitem{Murayama:1993em} 
  H.~Murayama and T.~Yanagida,
  Phys.\ Lett.\ B {\bf 322}, 349 (1994)
  [hep-ph/9310297].

\bibitem{Farhi:2007wj} 
  E.~Farhi, N.~Graham, A.~H.~Guth, N.~Iqbal, R.~R.~Rosales and N.~Stamatopoulos,
  Phys.\ Rev.\ D {\bf 77}, 085019 (2008)
  [arXiv:0712.3034 [hep-th]].

\bibitem{Gleiser:1993pt} 
  M.~Gleiser,
  Phys.\ Rev.\ D {\bf 49}, 2978 (1994)
  [hep-ph/9308279].

\bibitem{Landau:1976}
L.D. Landau and L. Lifshits, Mechanics (Pergamon Press, New York, 1976)

\bibitem{Fodor:2009kf} 
  G.~Fodor, P.~Forgacs, Z.~Horvath and M.~Mezei,
  Phys.\ Lett.\ B {\bf 674}, 319 (2009)
  [arXiv:0903.0953 [hep-th]].
  
\bibitem{Dashen:1975} 
  R.~F.~Dashen, B.~Hasslacher and A.~Neveu,
  Phys.\ Rev.\ D {\bf 11}, 3424 (1975).
    
\bibitem{Gradshteyn:1995}
Gradshteyn I.M Ryzhik, Table of Integrals, Series and Products, 4th ed.,
Academic, New York, 1965.

\bibitem{Kofman:1994rk} 
  L.~Kofman, A.~D.~Linde and A.~A.~Starobinsky,
  Phys.\ Rev.\ Lett.\  {\bf 73}, 3195 (1994)
  [hep-th/9405187].
  
\bibitem{forthcoming}
  M.~Kawasaki, N.~Takeda and K.~Nakayama, in preparation.
  
\end{thebibliography}
\end{document}